\documentclass[twocolumn]{revtex4}
\usepackage[colorlinks,bookmarksopen,bookmarksnumbered,citecolor=magenta,urlcolor=red]{hyperref}
\usepackage{graphicx}
\usepackage{amsmath}
\usepackage{natbib}
\begin{document}


\title{Tropical Vorticity Forcing and Superrotation in the Spherical Shallow Water Equations}

\author{D.L. Suhas, Jai Sukhatme and Joy M.
Monteiro} 
        
\begin{abstract}
The response of the nonlinear shallow water equations (SWE) on a sphere to tropical vorticity forcing is examined with
an emphasis on momentum fluxes and the emergence of a superrotating (SR) state.
Fixing the radiative damping and momentum drag timescales to be of the order of a few days,
a state of SR is shown to emerge under steady large-scale and random small-scale vorticity forcing.
In the first example, the stationary response to a pair of equal and oppositely signed vortices placed on the equator is considered.
Here, the equatorial flux budget is dominated by the eddy fluxes and these drive the system into a state of SR.
Eventually, the flux associated with the momentum drag
increases to balance the eddy fluxes, resulting in a steady state with a SR jet at the equator. The initial value problem with these twin vortices also exhibits SR driven by eddy fluxes. 
Curiously, this transient solution spontaneously propagates westward and continually circumnavigates the globe. It is worth emphasizing that 
this SR state does not rely on any particular form of dissipation at large scales, and is in fact observed even in the 
absence of any large-scale damping mechanism.
In the second example, a random small-scale vorticity forcing is applied across the tropics. The statistically steady state obtained is fairly 
turbulent in nature, but here too, the
eddy fluxes dominate, and the system exhibits SR.
It is important to note that in both these cases, the direct forcing of the zonal mean zonal flow is zero by construction, and the eddy fluxes
at the equator are responsible for its eastward acceleration.
Further, in both these examples, the rotational part of the flow dominates the momentum fluxes as well
as the stationary
zonal mean zonal flow itself.
Arguments based on the nature of potential vorticity 
and enstrophy are put forth to shed some light on these results.

\vskip 0.2truecm
\begin{center}
{\bf To appear in QJRMS, 2016.}
\end{center}

\end{abstract}
\vskip 0.25truecm


\maketitle

\section{Introduction}

Superrotation (SR) denotes a state of the atmosphere wherein the mean upper tropospheric tropical winds are westerly in nature \citep{held-sr}.
While these winds 
on present-day Earth have an easterly character \citep{po}, SR is a robust feature of other planetary atmospheres such as Venus 
\citep{gierasch}, Jupiter \citep{porco} and Saturn \citep{genio}. Though, it should be kept in mind that SR is not a completely extra-terrestrial 
phenomenon. Indeed, on Earth, the stratosphere \citep[in the westerly phase of the quasi-biennial oscillation,][]{held-sr, baldwin, dunkerton} as well as the thermosphere 
\citep{Ris} exhibit SR. 

Starting with two-level model studies \citep{hs,sd,sar}, SR has been examined in progressively detailed three dimensional, 
rotating stratified systems. In general, results from idealized simulations (i.e., without a seasonal cycle or stationary tropical 
waves) suggest that
the eddy momentum fluxes
required for SR arise from a competition between those from equatorial and 
extratropical Rossby waves; with the former (usually resulting from tropical convection) supporting, and the latter 
(generated primarily by baroclinic instability) inhibiting SR \citep[see for example,][and the references therein]{ls}. 
The inclusion of tropical stationary waves \citep{kh-2005} or enhanced moist convective activity 
\citep[say in a warmer world,][]{caba,arn} 
point to other possible means of generating SR. There is also evidence that on slowly rotating, shallow atmospheres SR
comes about by means of an unstable Rossby-Kelvin coupled mode that fluxes momentum into the tropics \citep{m1,m2,m3}.
In the context of present-day Earth, examining reanalysis data, or by simulations with explicit seasonal cycles, it is seen that, 
the momentum flux due to seasonal
changes in the meridional circulation plays an important role in suppressing SR in the troposphere \citep{lee,dwk,kh-2005}. 

In this work, we focus on SR in the shallow water equations (SWE). 
Recently SR has been seen to emerge in idealized forced-dissipative simulations of the single layer spherical SWE
\citep{sp}. In particular, \cite{sp} noted that SR is a robust feature of the circulation under global, small-scale, 
random forcing of the momentum equations, but only if the large scale sink takes the form of radiative damping. When 
momentum drag was introduced along with radiative damping, 
the statistically steady state reverted to one with extratropical
westerly jets and an easterly tropical flow. Continuing on this theme, \cite{si} and \cite{wd} confirmed the existence
of SR in the SWE, and highlighted the effects of radiative damping on the tilt of global Hough modes and on the momentum flux due to 
equatorially trapped Rossby waves in generating SR, respectively. Their analysis and numerical simulations were also performed with 
global random small-scale forcing of the momentum equations.

From the viewpoint of the SWE, possibly the simplest setup where one may expect SR is the classic Matsuno-Gill problem
\citep{mat,gill,wu}. Specifically, working within the context of an equatorial $\beta$-plane, 
\cite{mat} considered the steady response of the SWE to a periodic set of mass (heat) sources and sinks
distributed on the equator, while \cite{gill} examined the response to a single isolated source. In both cases, there was no
a priori mean background flow present. The solutions obtained in these two cases were tropically localized and their
rotational part consisted of Rossby wave gyres to the west of the source, straddling the equator. The Kelvin wave response to the east of the source was
qualitatively different in the two studies, with the \cite{gill} solution being more
pronounced and elongated, while that in \cite{mat} was understandably muted
given the presence of a neighbouring sink. Focusing on the rotational component of the solution, especially examining Figure 9 in \cite{mat},
the stationary Rossby waves can be seen to have a tilt in the North-West (NW) - South-East (SE) direction north of the equator and in SW-NE direction south of the equator. Considering that Rossby waves are the primary means of transporting energy in the system, this tilt is consistent, and indeed expected, from a Rossby wave source in the tropics. In
particular, the transport of energy away from the equator by Rossby waves results in an eddy momentum flux towards the equatorial region, and implies a
NW-SE tilt \citep{shp1}\footnote{It should be noted here that the tilt is dependent on the  drag and damping timescales ($\tau_m$ and $\tau_r$ respectively)
used in the momentum and mass equations. Specifically, from an extensive study of the $\tau_m,\tau_r$ parameter space, \cite{shp1} showed that, in 
general,
for moderate values (i.e, both of the order of a few days), the tilt is NW-SE. But with different timescales (especially $\tau_m \!\gg\! \tau_r$),
by reducing the
effective deformation scale, it is possible to obtain NE-SW tilting stationary solutions.}.

This convergence of eddy momentum flux into the tropics immediately raises the possibility
of SR in the Matsuno-Gill model. As it turns out, a mass source/sink configuration provides a
compensatory westward momentum flux that exactly balances the eddy flux over the equator and prevents the development of a SR state \citep{shp}.
Viewing the SWE as a representation of the upper troposphere, \cite{shp,shp1} then considered a modified form of the equations that included
direct forcing of the winds as a representation of the momentum transfer from the lower to upper troposphere. This set immediately leads to
a robust state of SR across a wide range of drag and damping timescales and forms an elegant route to explaining SR on other (especially tidally locked)
planets.

Here, we pursue other simple configurations wherein the spherical SWE can possibly exhibit SR.
In particular, we do not consider scenarios where the tropical forcing accelerates the zonal mean zonal winds; instead, our motivation is to 
construct a simple example where the resulting eddy fluxes lead to SR.
In the first example, we consider a set of two vortices, of equal and opposite strength, placed on the
equator. By construction, the direct momentum forcing is zero at the equator and the eddy fluxes dominate the zonal momentum balance, driving the system to a state of SR.
Eventually, the momentum flux associated with the drag terms balances the eddy momentum flux at the equator, leading to a stationary solution. 
Away from the equator, the eddy, vorticity and momentum source fluxes largely cancel each other with the residual again balanced by momentum drag.
Furthermore, we see a fair degree of linearity in the system. Specifically,
the solution with a source and sink of vorticity is very well approximated by the superposition of stationary solutions to
a single localized
vorticity source and sink, respectively. We also consider the initial value problem with this configuration of twin vortices on the equator. 
The solution maintains its coherence for the length of the simulation, and is characterised by a quadrupole structure localised in the subtropics. In addition, the transient
solution spontaneously propagates westward and circumnavigates the globe. In the absence of damping
or drag, the quadrupole structure displays no discernible change in its amplitude for the length of the simulation.

The second example looks at random small-scale forcing applied across the tropics. In this scenario as well, the eddy fluxes dominate and push the system
to a SR state. This example is more turbulent in nature in that the stationary state is characterized by power at all scales.  In fact, the kinetic energy displays a -5/3 scaling in its power spectrum. In both cases, the stationary solution and fluxes are dominated by rotational contributions.
Finally,
we attempt to interpret these results by using the
conservation (or lack thereof) of potential vorticity (PV) and enstrophy.

\section{Basic Equations}

The shallow water equations with a mass and momentum forcing, radiative damping and momentum drag take the form,
\begin{align}
\frac{Du}{Dt} - fv = -g \frac{\partial h}{\partial x} - \frac{u}{\tau_m} + F_u, \nonumber \\
\frac{Dv}{Dt} + fu = -g \frac{\partial h}{\partial y} - \frac{v}{\tau_m} + F_v, \nonumber \\
\frac{Dh}{Dt} + h (\nabla \!\cdot \!{\bf u}) =  -\frac{(h-H)}{\tau_r} + S.
\label{1}
\end{align}
Here $\frac{D}{Dt} = \frac{\partial}{\partial t} + {\bf u}\!\cdot\!\nabla $, ${\bf u}=(u,v)$ is the horizontal flow, 
$f$ is the Coriolis parameter and $h(x,y,t)$ is the depth of the fluid
($H$ is the mean undisturbed depth, taken to be a constant). The drag in the momentum equations has the same timescale in the zonal
and meridional directions ($\tau_m$), $\tau_r$ is the radiative damping timescale, $S$ is the mass forcing and $F_u,F_v$ are the
momentum source terms.

Decomposing all fields into a zonal mean and departure therefrom, denoted by $\overline{()}$ and $()'$ respectively, and using the flux form of the  momentum equations, we obtain \citep{thuburn,shp},
\begin{align}
\frac{\partial \overline{u}}{\partial t} -  \underbrace{\overline{v}^* \overline{\omega_a}}_\text{vorticity} = \underbrace{-\frac{1}{\overline{h}} \partial_y [\overline{h}~\overline{u'v'}+ \overline{v}\overline{u'h'}]}_\text{eddy} 
- \underbrace{\frac{\partial_t \overline{h'u'}}{\overline{h}}}_\text{transient}
- \underbrace{\frac{\overline{u'h'}}{\overline{h}\tau_r}}_\text{radiative} \nonumber \\
+ \underbrace{\frac{\overline{S}~\overline{u}}{\overline{h}} + \frac{\overline{S'u'}}{{\overline{h}}}}_\text{mass forcing}
- \underbrace{\frac{\overline{u}}{\tau_m} - \frac{\overline{u'h'}}{\overline{h} \tau_m}}_\text{momentum drag} 
+ \underbrace{\overline{F_u} + \frac{\overline{h'F_u'}}{\overline{h}}}_\text{momentum source},
\label{2}
\end{align}
where $\overline{v}^* = \overline{vh}/\overline{h}$ and $\omega_a=\omega + f$ is the absolute vorticity. $\partial_t$ and $\partial_y$ refer to 
partial derivatives with respect to time and $y$, respectively. The second term on the LHS of (\ref{2}) is
called the vorticity flux. On the RHS, the first term is the eddy momentum flux, the second is a transient flux, the third term is referred to as
the radiative flux,
the fourth and fifth terms are the fluxes due to mass forcing, sixth and seventh terms are the fluxes due to momentum drag, and the last two terms are the fluxes
associated with the momentum source. Together, these fluxes accelerate the
zonal mean zonal flow \citep{thuburn}.  In terms of vorticity fluxes alone, another form of the evolution equation for the zonal mean zonal flow ($\overline{u}$), derived
by taking a zonal mean of the zonal momentum equation reads,
\begin{equation}
\frac{\partial \overline{u}}{\partial t} - \overline{v} ~\overline{\omega_a} - \overline{v'\omega'} = - \frac{\overline{u}}{\tau_m} + \overline{F_u}.
\label{2a}
\end{equation}
Defining the potential vorticity (PV), $Q$, as $\omega_a/h$, its evolution is given by,

\begin{equation}
\frac{\partial Q}{\partial t} + ({\bf{u}}.\nabla) Q = \frac{Q (h-H)}{h \tau_r} - \frac{SQ}{h} - \frac{\omega}{h \tau_m} + \frac{(\partial_x F_v - \partial_y F_u)}{h},
\label{2b}
\end{equation}

which shows that a radiative sink can act as a source for PV \citep[see for example,][]{momc}. Also,
an equation for PV-substance (i.e., $hQ$) reads,
\begin{equation}
\frac{\partial (hQ)}{\partial t} + \nabla \!\cdot \!(hQ{\bf u}) = -\frac{\omega}{\tau_m} + (\partial_x F_v - \partial_y F_u).
\label{3}
\end{equation}
Thus, in the absence of momentum drag and momentum forcing, $hQ$ obeys a conservation law \citep{hm1,hm2}. But, here, $hQ=\omega_a$, therefore
(\ref{3}) becomes,
\begin{equation}
\frac{\partial \omega}{\partial t} + \nabla \!\cdot \!(\omega_a{\bf u}) = -\frac{\omega}{\tau_m} + (\partial_x F_v - \partial_y F_u).
\label{4}
\end{equation}
Indeed, a zonal mean of (\ref{4}) also yields (\ref{2a}). 

\subsection{Superrotation}

Here, following \cite{shp}, we list the conditions for SR. But, rather than just asking for westerly winds at the equator, we use a slightly more 
restrictive definition of SR, namely, we require $\overline{u}$ to be
positive and have
a local maximum at the equator. Consider (\ref{2a}), 
in  steady state, at
the equator (i.e., $\omega_a=\omega$).
\begin{equation}
 \overline{v} ~\overline{\omega} + \overline{v'\omega'} + \overline{F_u} - \frac{\overline{u}}{\tau_m} = 0.
\label{a1}
\end{equation}
But, if the zonal mean zonal wind has an extremum at the equator, then $\partial_y \overline{u} = 0$. Further, $\overline{\omega} = -\partial_y \overline{u}$,
so, at the equator $\overline{\omega}=0$. Using this in (\ref{a1}),
\begin{equation}
 \frac{\overline{u}}{\tau_m} = \overline{v'\omega'} + \overline{F_u}.
\label{a2}
\end{equation}
Thus, a positive extremum of $\overline{u}$ at the equator can be maintained by $\overline{v'\omega'}$ or directly by the
forcing via $\overline{F_u}$.  As mentioned, the example provided by \cite{shp1} involves a non-zero $\overline{F_u}$ which
in turn drives their model to a state of SR. In this work, we impose $\overline{F_u}=0$, i.e., we try to construct examples wherein
the eastward acceleration is provided only by the eddy fluxes.

Keeping this in mind, we now look at the enstrophy, i.e., we take (\ref{4}), decompose all fields into a zonal average and
deviation, multiply by $\omega'$ and perform a zonal average. This yields (at the equator),
\begin{equation}
\overline{v'\omega'} \frac{\partial \overline{\omega}}{\partial y} + \overline{(\omega')^2} \frac{\partial \overline{v}}{\partial y}
+ \overline{v} ~\overline{\omega' \partial_y \omega'} = -\frac{\overline{(\omega')^2}}{\tau_m} + \overline{\omega' F_\omega'},
\label{a3}
\end{equation}
here, $F_\omega' = \partial_x F_v' - \partial_y F_u'$.  Furthermore, we have assumed a steady state, neglected third order terms and used $\overline{\omega}=0$. 
A more symmetric form of (\ref{a3}) is,
\begin{equation}
\overline{v'\omega'}~ \frac{\partial \overline{\omega}}{\partial y} + \frac{1}{2} (\overline{{\bf u}} \!\cdot\! \nabla) \overline{(\omega')^2}
+ \overline{(\omega')^2} ~[\nabla \!\cdot\! \overline{{\bf u}} ] = \overline{\omega' F_\omega'} -  \frac{\overline{(\omega')^2}}{\tau_m}.
\label{a5}
\end{equation}
Its easy to see from (\ref{a5}) that if the zonal mean state is incompressible then the second and third terms on the LHS are zero.
As $\overline{F_u}=0$, for a positive $\overline{u}$ we require $\overline{v'\omega'}$ to also be
positive from (\ref{a2}). 
Further, for a local maximum of $\overline{u}$ at the equator, $\partial_y \overline{\omega} = -\partial_{yy} \overline{u}$
has to be positive. If the second and third terms on the LHS of (\ref{a5}) are zero then the LHS is positive, the only way this is possible is if
$\overline{\omega' F_\omega'}$ is non-zero and positive (as the other term on the RHS is negative). Therefore, if the mean flow is incompressible, 
even with $\overline{F_u}=0$, it is possible
to obtain SR via a positive $\overline{\omega' F_\omega'}$.
In fact, 
a forcing that correlates with the eddy vorticity is required to support our
notion of SR. Indeed, this is essentially the argument presented by \cite{shp1}, though they explicitly work in the 
quasigeostrophic framework.

Thus, forcing can directly lead to this slightly stricter form of SR if $\overline{F_u} \neq 0$ or, if $\overline{F_u}=0$, by positive eddy contribution from 
$\overline{v' \omega'}$ provided that the stationary state is mainly rotational in character.
Note further that a mass forcing cannot do this as the $S$ term does not enter (\ref{a2}) or (\ref{a5}), and
will not contribute to $\overline{v'\omega'}$, nor will it be able to directly support a positive maximum of $\overline{u}$ at the 
equator. Indeed, this is likely the reason why the Matsuno-Gill problem does not
exhibit SR \citep{shp}.

\section{Numerical Setup}

The shallow water system is solved in spherical geometry using the SHTns library \citep{shtns}.
All results are reported at a resolution of 512 (longitude) $\times$ 256 (latitude), triangularly trucated corresponding to a maximum resolved wavenumber of 170.
We use a 3$^{rd}$ order Adams-Bashforth integrator for time stepping. A $\triangle ^4$ hyperviscosity is used for small scale dissipation.
The mean depth of the fluid is fixed at 300 m\footnote{Experiments with different mean depths
were also performed, and the results obtained do not change the conclusions presented here.}. Planetary radius ($a$), rotation rate ($\Omega$) and acceleration due to gravity ($g$) are set to that of the Earth.
This yields $L_D/a = 0.06$ and $L_{eq}/a = 0.24$, where $L_D = \sqrt{gH}/(2 \Omega)$ is the polar deformation radius 
and $L_{eq} = \sqrt{a L_D}$ is the equatorial deformation radius \citep{sp2}.
A range of radiative and momentum drag scales were
considered, but for the most part we set $\tau_r \sim \tau_m = 10$ days.

\section{Large-Scale Forcing: Stationary Solution}

We follow an analog of the \cite{mat} problem, i.e., the response of the SWE to periodic, equatorial, large-scale vorticity forcing. In particular,
we consider two vortices, one positive and one negative of the form,
\begin{align}
F_\omega = F_0  \exp[-((\phi - \phi_0) / \triangle \phi)^2] \{ \exp[-((\lambda - \lambda_1) / \triangle \lambda)^2] \nonumber \\
- \exp[-((\lambda - \lambda_2) / \triangle \lambda)^2] \} ,
\label{eqn_fr}
\end{align}
where $F_0$ is the amplitude of the forcing, $\triangle \phi = 10^\circ$ and $\triangle \lambda = 30^\circ$ are the latitudinal and longitudinal half-width, 
respectively. Here, $\phi_0 = 0$, $\lambda_1 = 90^\circ$ and $\lambda_2 = 270^\circ$. 
The choice of $\triangle \phi = 10^\circ$ ensures that the forcing is well within the equatorial
deformation scale. Further, the choice of forcing sets $\overline{F_u}=0$, and the steady state balance in (\ref{a2}) is between the 
drag and $\overline{v' \omega'}$. Finally, as there is no mass forcing, $S=0$ in our examples.

\begin{figure}[]
\centering
\includegraphics[width=8cm]{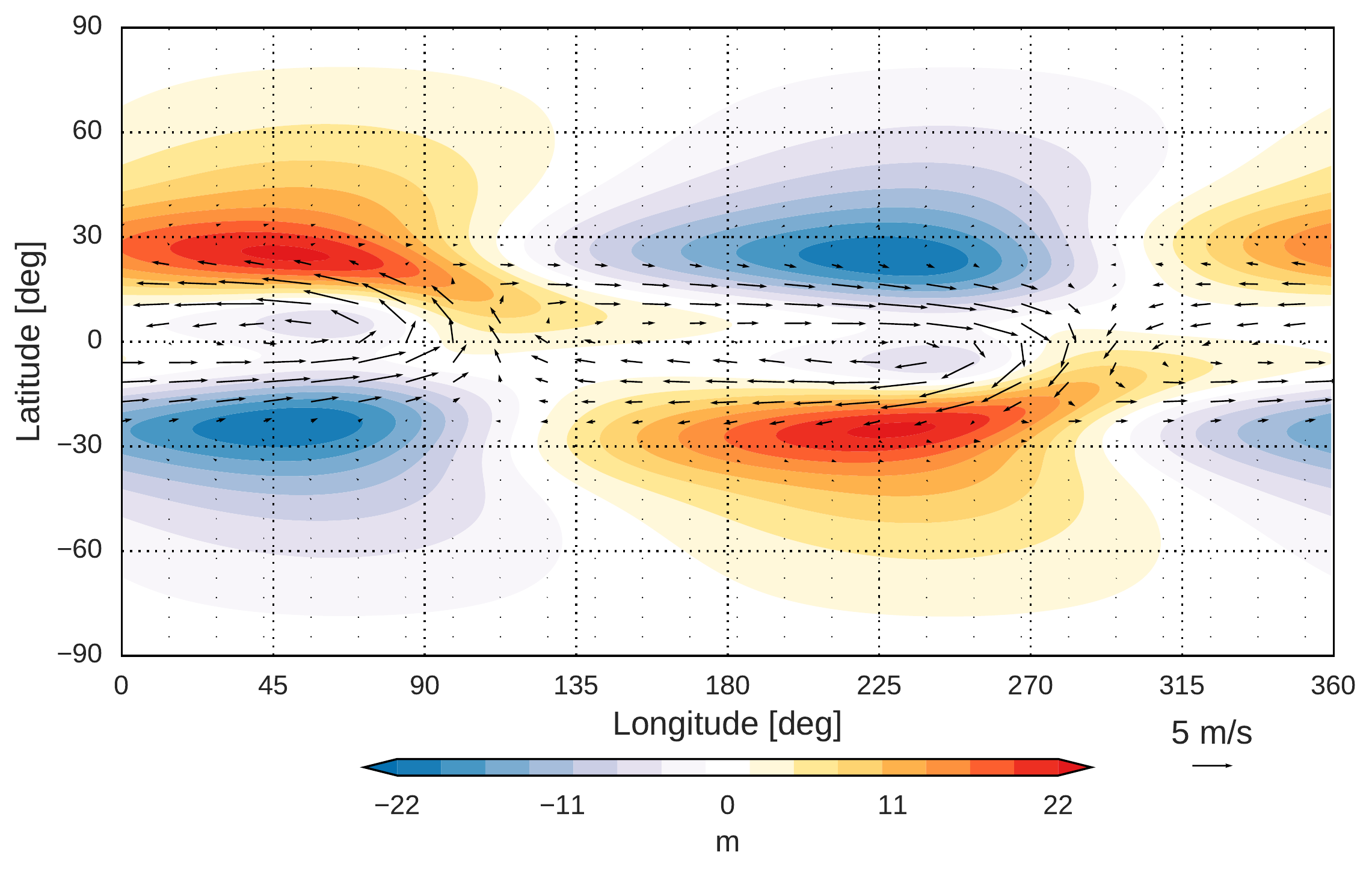}
\caption{\label{fig_height} Stationary height perturbations for the vorticity source and sink placed on the equator, along with velocity perturbation quivers.}
\end{figure}

\begin{figure*}[]
\centering
\includegraphics[width=15cm]{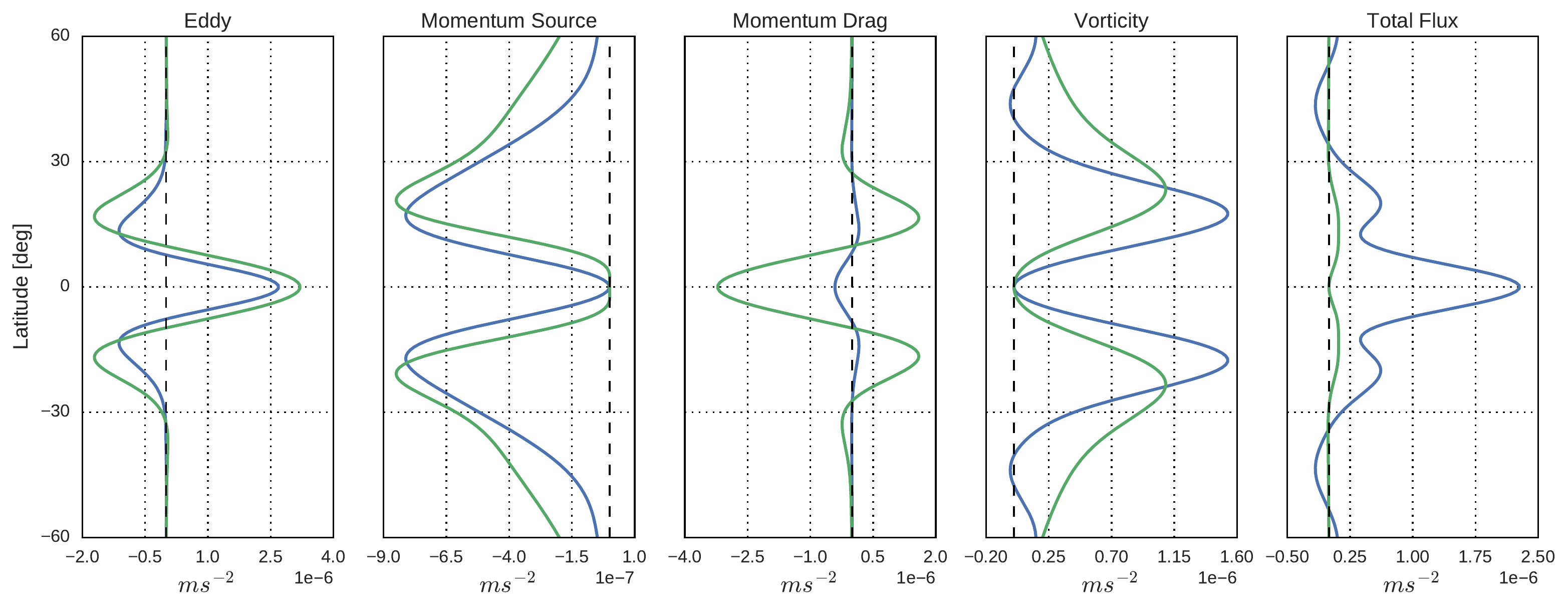}
\caption{\label{fig_flux} Momentum fluxes for the vorticity source and sink placed on the equator. The panels show the eddy, momentum source, momentum drag, vorticity and total fluxes as defined in Equation \ref{2}. The blue curves are for day 2 
and green curves for day 100 (steady state). Black dashed line denotes the zero flux.}
\end{figure*}

The stationary solution obtained with this forcing is shown in Figure \ref{fig_height}. Quite clearly, the solution 
has a strong subtropical signature consisting of a distinct quadrupole structure. In the northern hemisphere, the height field shows an anticyclone immediately 
polewards of the source, east of
which is a cyclonic eddy. 
The anticyclonic eddies show an equatorward extension that is missing from their cyclonic counterparts.
Further, on the equatorward side, there is a slight tilt of eddies (clearer for the anticyclones) that is oriented in a NW-SE manner, suggestive of an eddy
momentum flux towards the equator. 
The momentum fluxes at early and late times, 
associated with this forcing,
are shown in Figure \ref{fig_flux} 
(the momentum source flux here consists entirely of $\overline{h' F_u'}/\overline{h}$, we have not shown the fluxes associated with the time 
derivative and radiative terms as these are more than an order of magnitude 
smaller than the others).
At the equator, the eddy flux dominates.
Of course,
with time, the drag flux grows and balances the momentum budget at the equator. Note that off the equator, in the subtropics, the vorticity flux
largely offsets the momentum source and eddy fluxes. Their difference is also balanced by the drag flux at long times.

\begin{figure}[]
\centering
\includegraphics[width=8cm]{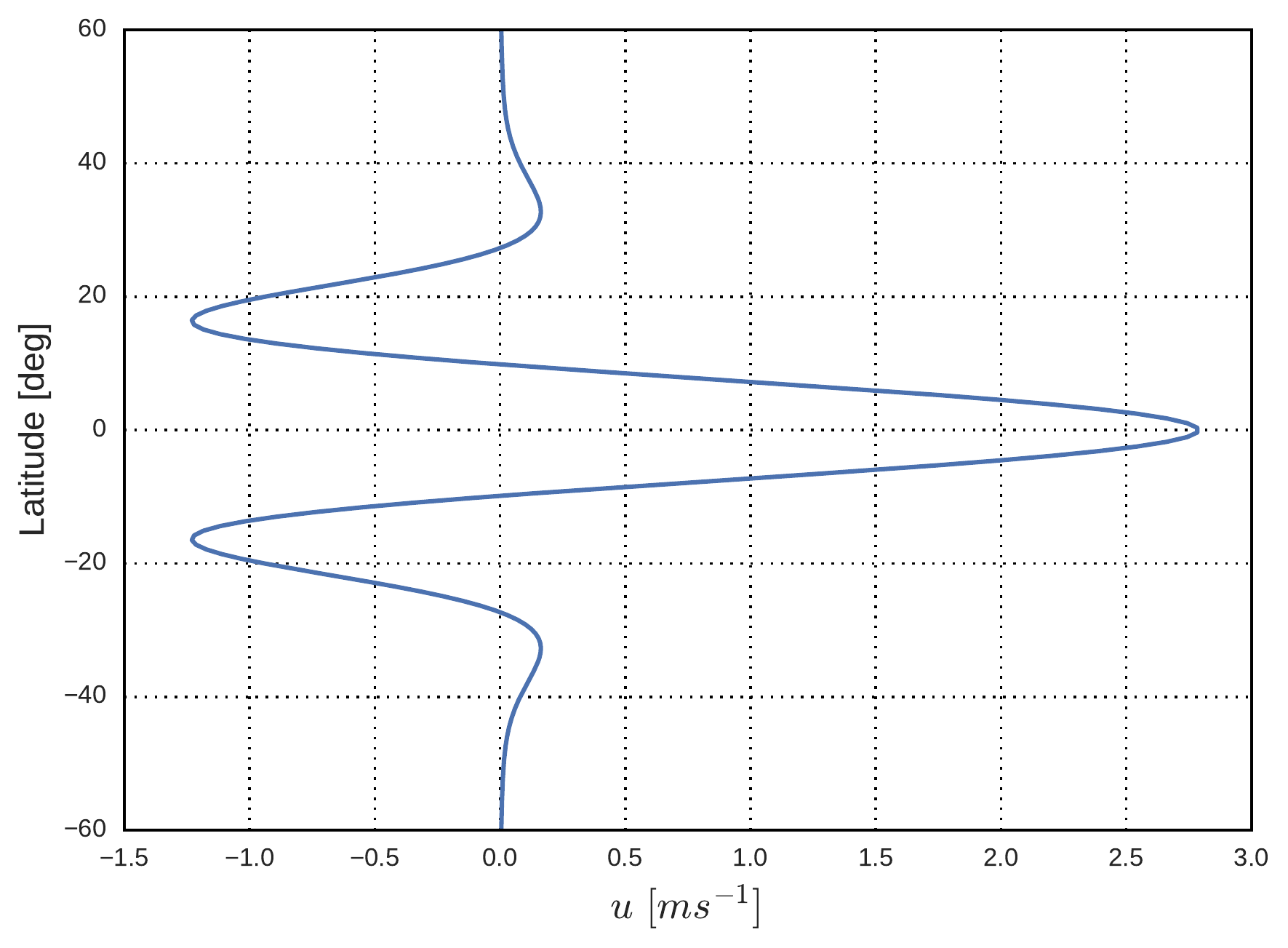}
\caption{\label{fig_uzon} Steady state zonal mean zonal velocity profile for a vorticity source and sink placed on the equator.}
\end{figure}

\begin{figure}[]
\centering
\includegraphics[width=8cm]{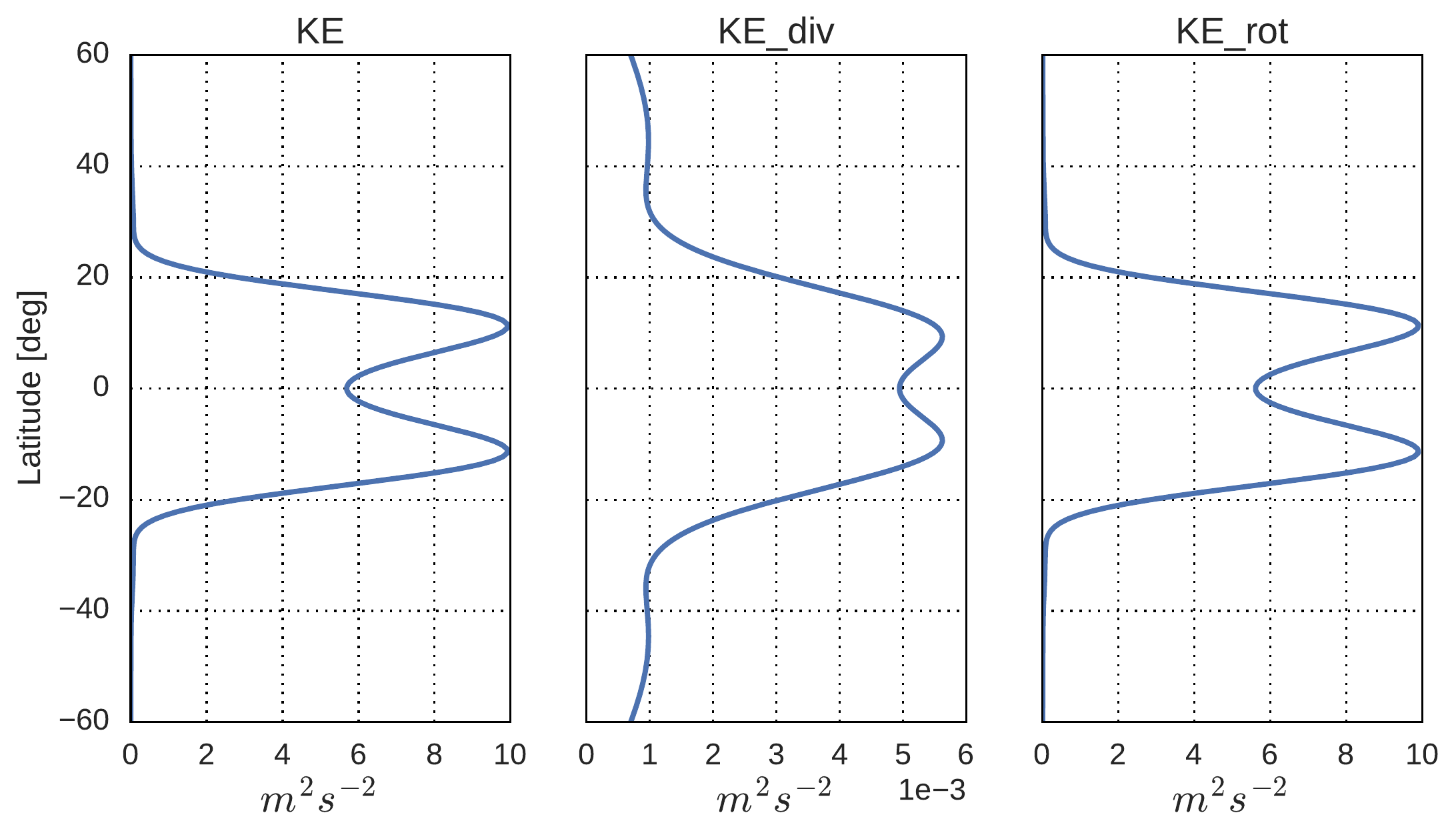}
\caption{\label{fig_kesplit} Zonal mean of steady state kinetic energy and its divergent and rotational  components for a vorticity source and sink placed on the equator.}
\end{figure}

Consequently, as is seen in Figure \ref{fig_uzon}, the zonal mean zonal flow has a clear positive maximum at the equator, and the system is in a state of 
SR (even by our slightly restrictive definition).
Figure \ref{fig_kesplit} shows the zonal mean kinetic energy and its divergent and rotational components. 
Quite clearly, the divergent part is negligible, and the entire zonal mean flow is rotational in nature. This gives a somewhat after-the-fact credence to the argument 
surrounding (\ref{a5}), specifically,
that the state of SR is linked to vorticity forcing and the implied positive $\overline{\omega' F_\omega'}$ term. Indeed, we have checked that in a
stationary state, not only is $\overline{\omega' F_\omega'}$ positive, there is an accurate balance between the $\overline{v' \omega'}$ and 
$\overline{u}/\tau_m$ at the equator.

\begin{figure*}[]
\centering
\includegraphics[width=15cm]{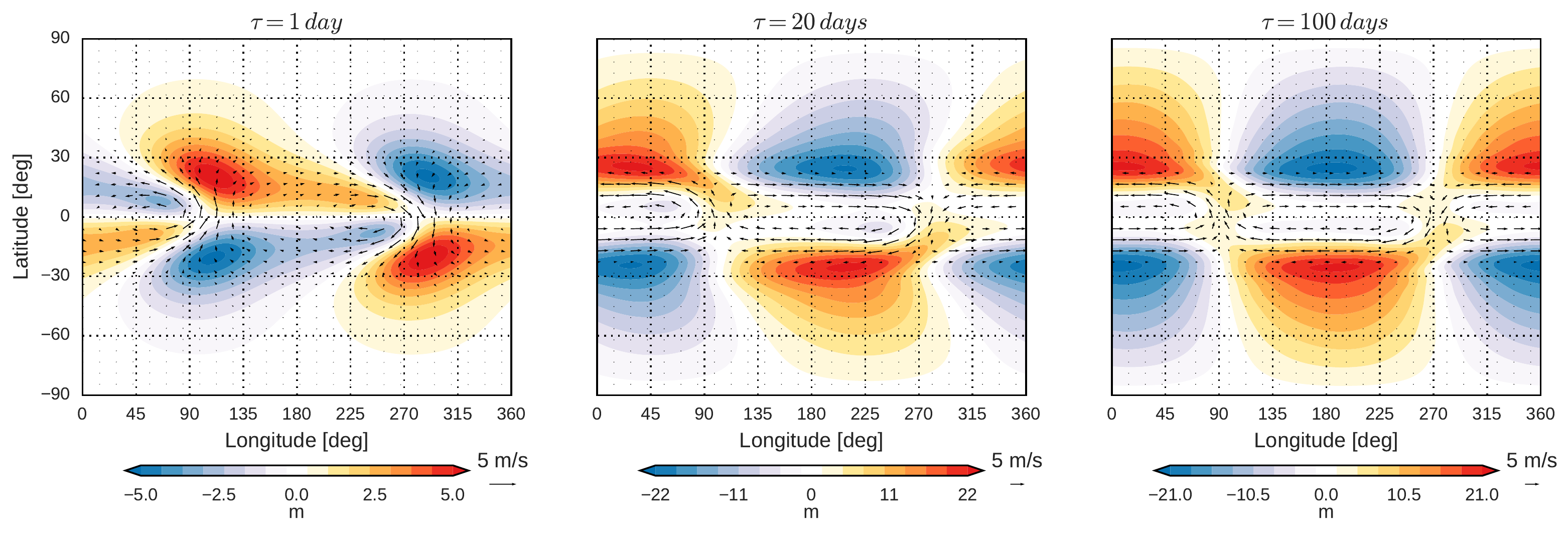}
\caption{\label{fig_height_tau} Steady state height perturbations for the vorticity source and sink placed on the equator along with velocity perturbation quivers for different drag timescales.}
\end{figure*}

\begin{figure*}[]
\centering
\includegraphics[width=15cm]{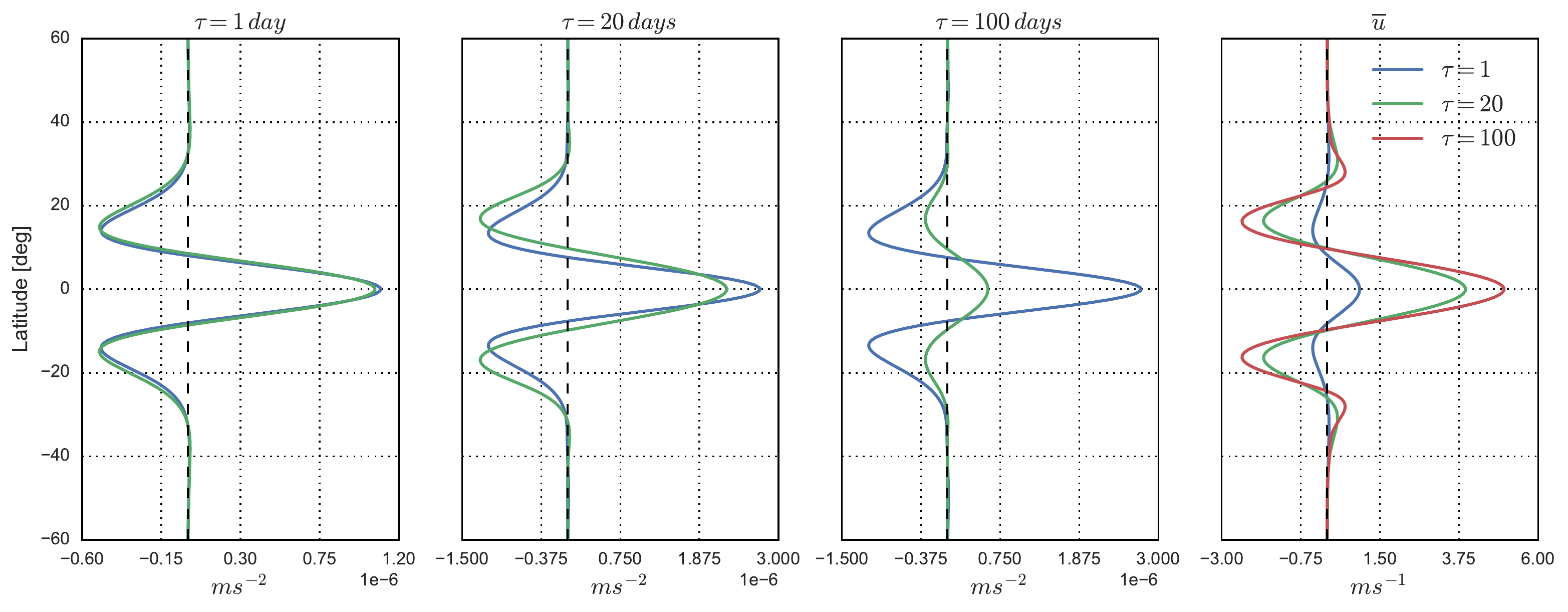}
\caption{\label{fig_flux_tau} The first three panels show eddy fluxes with different drag and damping timescales. Here the 
initial and steady state fluxes are shown in blue and green, respectively. The final panel shows the zonally averaged 
zonal velocities for the three drag and damping timescales. For clarity $\overline{u}$ for $\tau=1$ curve has been multiplied by a factor of 10.  Black dashed line denotes the zero flux.}
\end{figure*}

The dependence of the stationary solution on the drag and damping timescales is examined in Figure \ref{fig_height_tau}. Here we show the stationary
solution for $\tau_r=\tau_m=1,20$ and $100$ days. 
The associated eddy fluxes and the zonal mean zonal flows for these runs are shown in Figure \ref{fig_flux_tau}. 
In all cases the zonal mean zonal flow is in a state of SR. Further, the emergence of SR is always tied to the eddy fluxes. Interestingly, 
the NW-SE tilt of the stationary
eddies is most pronounced for the strong drag and damping case, i.e., at $\tau_r=\tau_1=1$ day. Also, the tilt of the eddies at higher latitudes (poleward of $30^\circ$N/S, respectively) is oriented in a reverse NE-SW direction, a feature that most clear in the $\tau_r=\tau_m=20$ days case. 
Further, much like \cite{shp1}, the solution for high
drag and damping connects the equatorial and subtropical regions, while the tendency for a strong subtropical signature and further extension into
higher latitudes increases as these large scale 
sinks become weaker.

\begin{figure*}[]
\centering
\includegraphics[width=15cm]{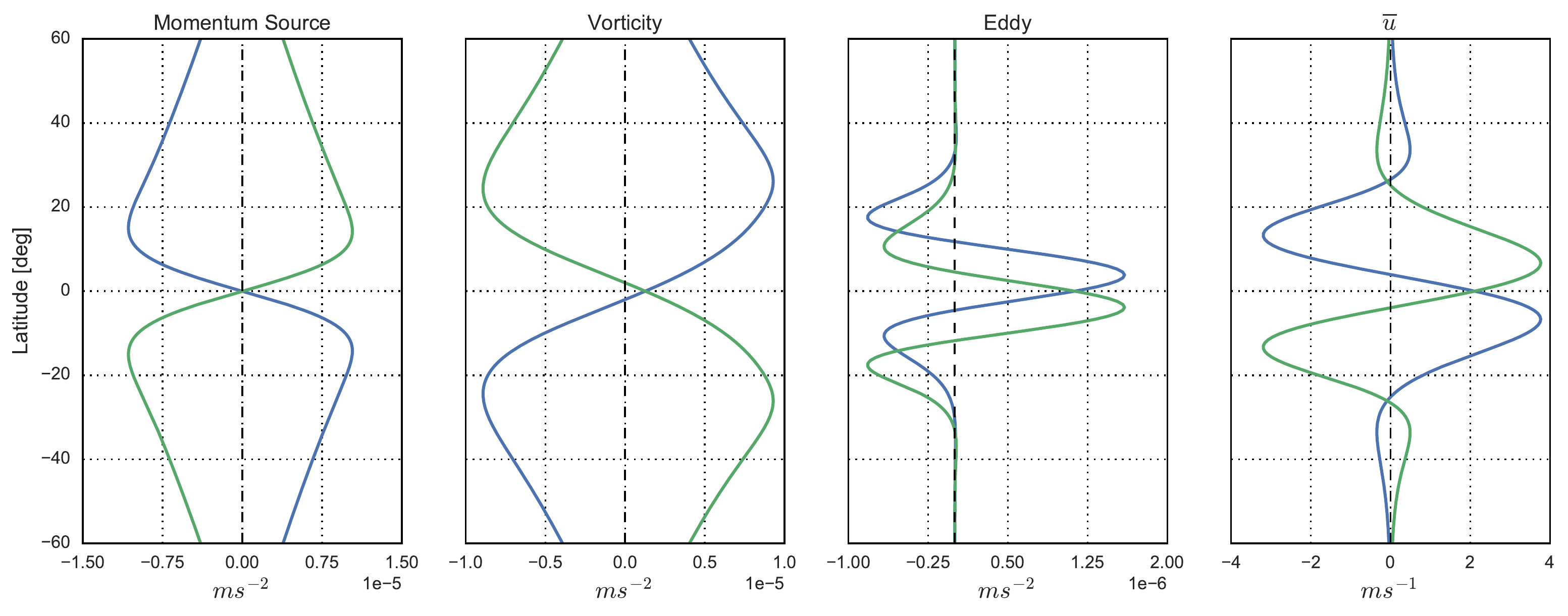}
\caption{\label{fig_flux_single_both} Momentum fluxes for single vortex at steady state. First three panels are the steady state momentum source, vorticity and eddy fluxes as defined in Equation \ref{2}. The final panel shows the zonally averaged zonal velocities. Blue (green) curve is for the vortex source (sink). Black dashed line denotes the zero flux.}
\end{figure*}

It proves instructive to decompose the forcing and examine the fluxes for a single positive and negative vortex at the equator. The momentum source, vorticity and
eddy fluxes along with the zonal mean zonal flow from these experiments are shown in Figure \ref{fig_flux_single_both}. 
We immediately note that the individual
vortices on the equator do not result in a clear state of SR. 
In fact, the vorticity and momentum source flux dominates to result in oppositely oriented jets on either side of the
equator\footnote{Note that the momentum source flux in this case comprises of both $\overline{F_u}$ and $\overline{h'F_u'}/\overline{h}$, in fact, it is dominated
by the former term. But here too, by construction,
$\overline{F_u}=0$ at the equator.}. 
Indeed, the eddy flux is still eastward in character but attains a maximum off the equator. Further, it is an order of magnitude smaller that the 
fluxes associated with the vorticity and momentum source terms. Note that the eddy flux is uncompensated at the equator and does lead to a small positive 
zonal mean zonal flow here, but is dwarfed in Figure \ref{fig_flux_single_both} by the jets off the equator 
(technically, it does satisfy the condition of westerlies at the equator but does not meet the
slightly stricter criterion for SR that we have used). 
More interestingly, comparing the fluxes of individual vortices (Figure \ref{fig_flux_single_both}) with those of 
the two vortices together (Figure \ref{fig_flux}), suggests linearity in that there appears to be a fair degree of superposition with these forcing
protocols. 

\section{Large-Scale Forcing: Initial Value Problem}
Having examined the stationary solution, it is worth studying the freely evolving problem with the aforementioned vorticity source and sink
as the initial condition. One of our motivations in doing so is to see if SR observed in a stationary state also emerges
when the dynamics is not subjected to continual forcing. Further, even though large-scale vortices can exist at lower latitudes (examples include
the long lived Great Red Spot on Jupiter that is located at approximately $22^\circ$S, as well as the more transient twin cyclones that form near 
the equator on Earth), 
we would like to see the fate of twin vortices, i.e.,
does the solution maintain its coherence or is it immediately distorted or torn into smaller scale features.

\begin{figure*}[]
\centering
\includegraphics[width=15cm]{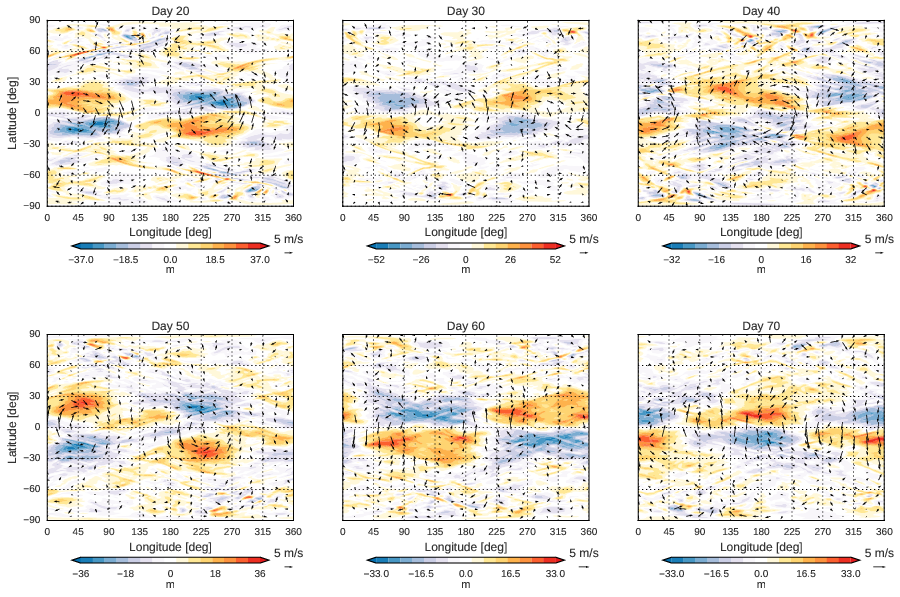}
\caption{\label{fig_height_ivp} Evolution of height perturbations for the initial value problem along with velocity perturbation quivers.}
\end{figure*}

\begin{figure}[]
\centering
\includegraphics[width=8cm]{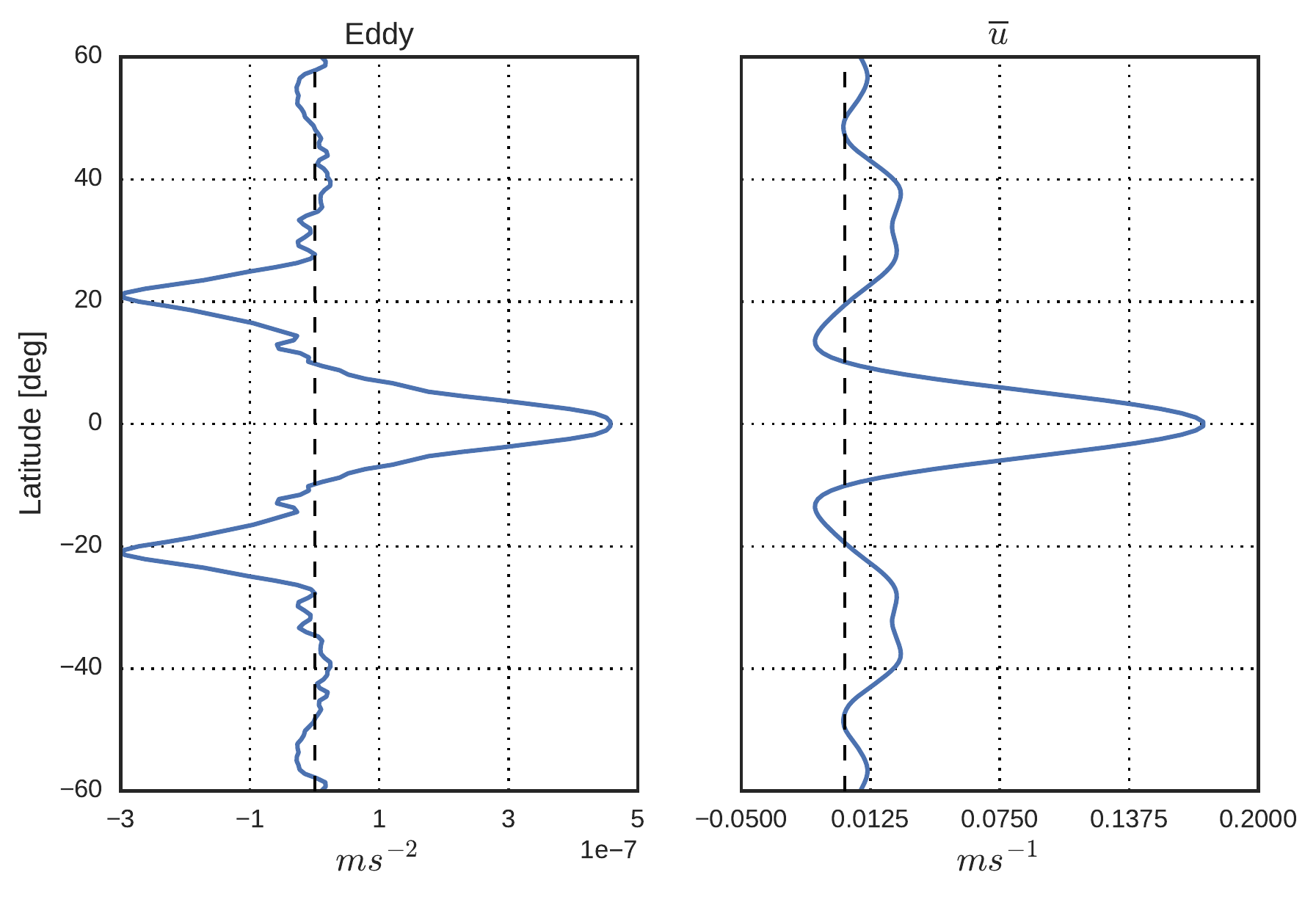}
\caption{\label{fig_flux_ivp} Eddy flux and zonally averaged zonal velocity for the initial value problem averaged over days 10 - 100. Black dashed line denotes the zero flux.}
\end{figure}

The initial field comprises of vorticity source and sink centred at the equator with similar configuration as used in large scale steady forcing (see Equation \ref{eqn_fr}). 
As is seen in Figure \ref{fig_height_ivp}, which depicts the height field and velocity perturbations for days 20-70, the quadrupole structure of the solution is 
established fairly quickly and persists through these days. Indeed, with no damping and drag (setting $\tau_m=\tau_r=\infty$), the solution maintains
its large-scale, predominantly subtropical signature throughout the simulation.
Associated with this, the zonal mean zonal flow is shown in 
Figure \ref{fig_flux_ivp}. Clearly, the system still exhibits SR\footnote{We have also carried out this experiment with $\tau_r=\tau_m=10$ days. SR as well as 
the basic structure of the quadrupole remains the same, only the entire field dies out in about 100 days.}. 
As this is a transient solution, it is somewhat difficult to estimate momentum fluxes. Even so, when averaged over a fairly long period
(from day 10-100), the eddy fluxes (also shown in Figure \ref{fig_flux_ivp}) show a distinct positive maximum at the equator 
and appear to be responsible for keeping the
system in a SR state.

A remarkable feature of the initial value problem is the spontaneous longitudinal propagation of the solution, specifically, returning once again
to Figure \ref{fig_height_ivp}, we see that the quadrupole moves westward with time. Indeed, the signal in the subtropical 
belt continually circumnavigates the globe. 
Furthermore, 
the system maintains the coherence of its cyclonic and anticyclonic
structures through time as can be seen by comparing the height fields at day 30 and day 60 in Figure \ref{fig_height_ivp}.

\section{Small-Scale Forcing}
Motivated by the results of \cite{sp} \citep[see also,][]{wd,si} we now consider random small-scale forcing of the momentum equations, but restrict
the forcing to the tropical belt (between 20$^{\circ}$ N and 20$^{\circ}$ S). In particular, the spectrum of forcing takes the form, $F_n = \hat{A}(1-r^2)e^{i\theta}+rF_{n-1}$,
where $\theta$ is a random number in $[0,2\pi)$, $\hat{A}$ is the wavelength-dependent forcing amplitude, $r$ is the correlation coefficient
and the subscript on $F$
denotes the timestep \citep{malval,sp2}. The forcing is restricted to  $\vert k - k_f \vert \leq 2 $, where $k$ is the total wavenumber. In all our simulations $r=0$ and $k_f=100$.
The radiative and momentum drag scales , $\tau_r$ and $\tau_m$ are set to 100 days.
 
\begin{figure}[]
\centering
\includegraphics[width=8cm]{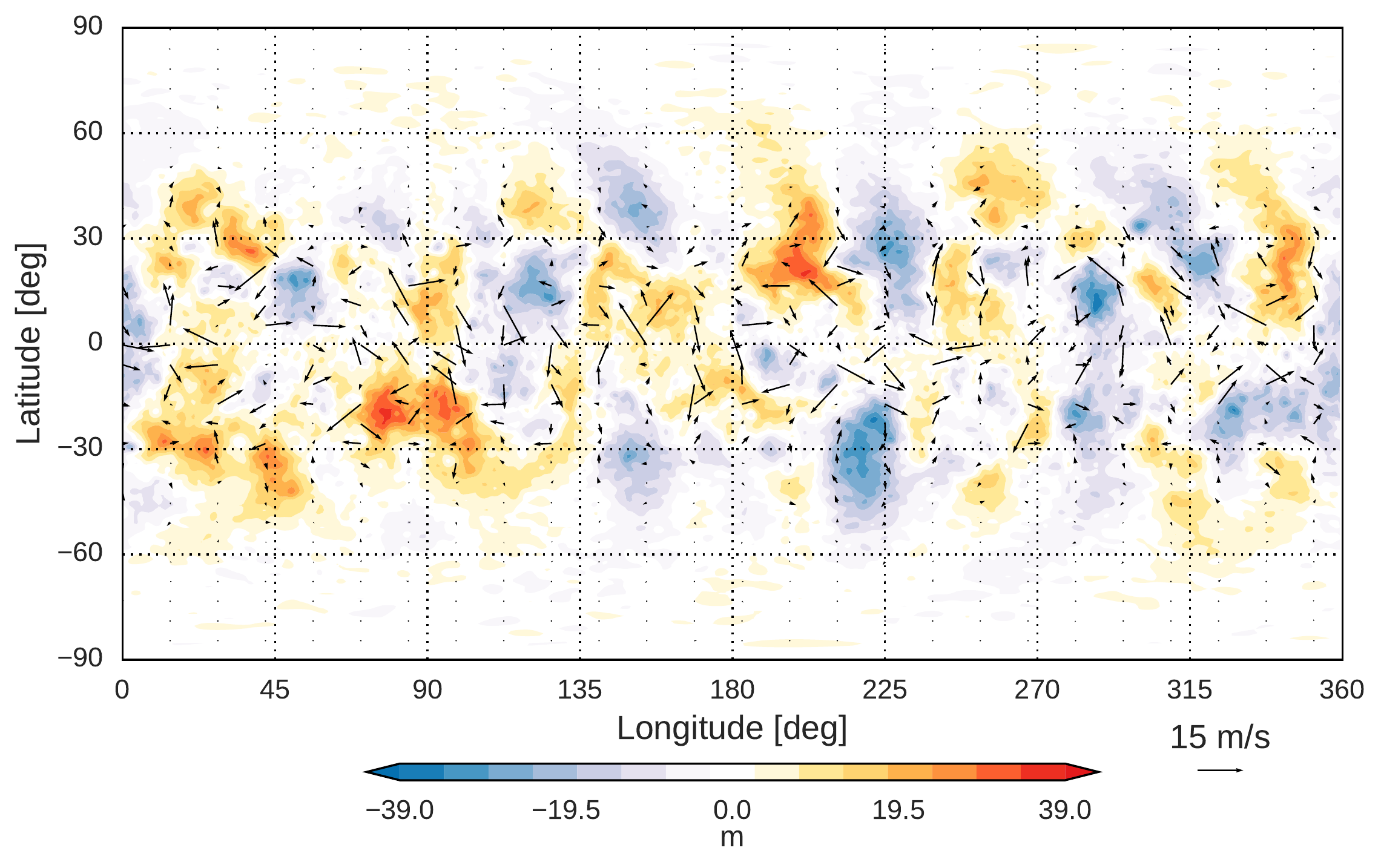}
\caption{\label{fig_height_turb} Snapshot of the height perturbations (after settling into a statistically steady state) 
for the small scale randomly forced scenario. }
\end{figure}

\begin{figure}[]
\centering
\includegraphics[width=8cm]{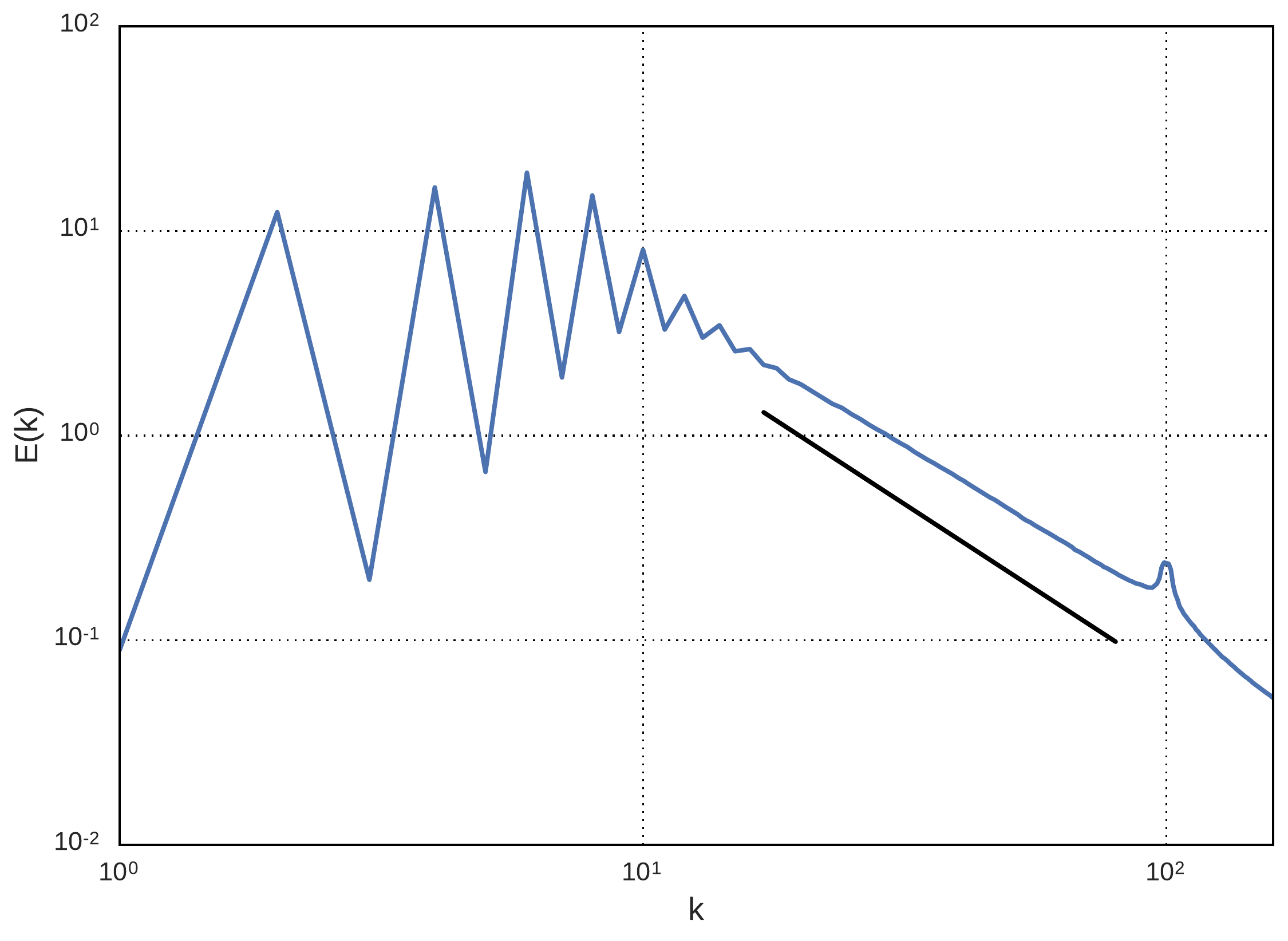}
\caption{\label{fig_spectra} Kinetic energy spectra for small-scale random forcing. The spectrum is averaged over days 500-2000. Black line has a slope of -5/3.}
\end{figure}

\begin{figure}[]
\centering
\includegraphics[width=8cm]{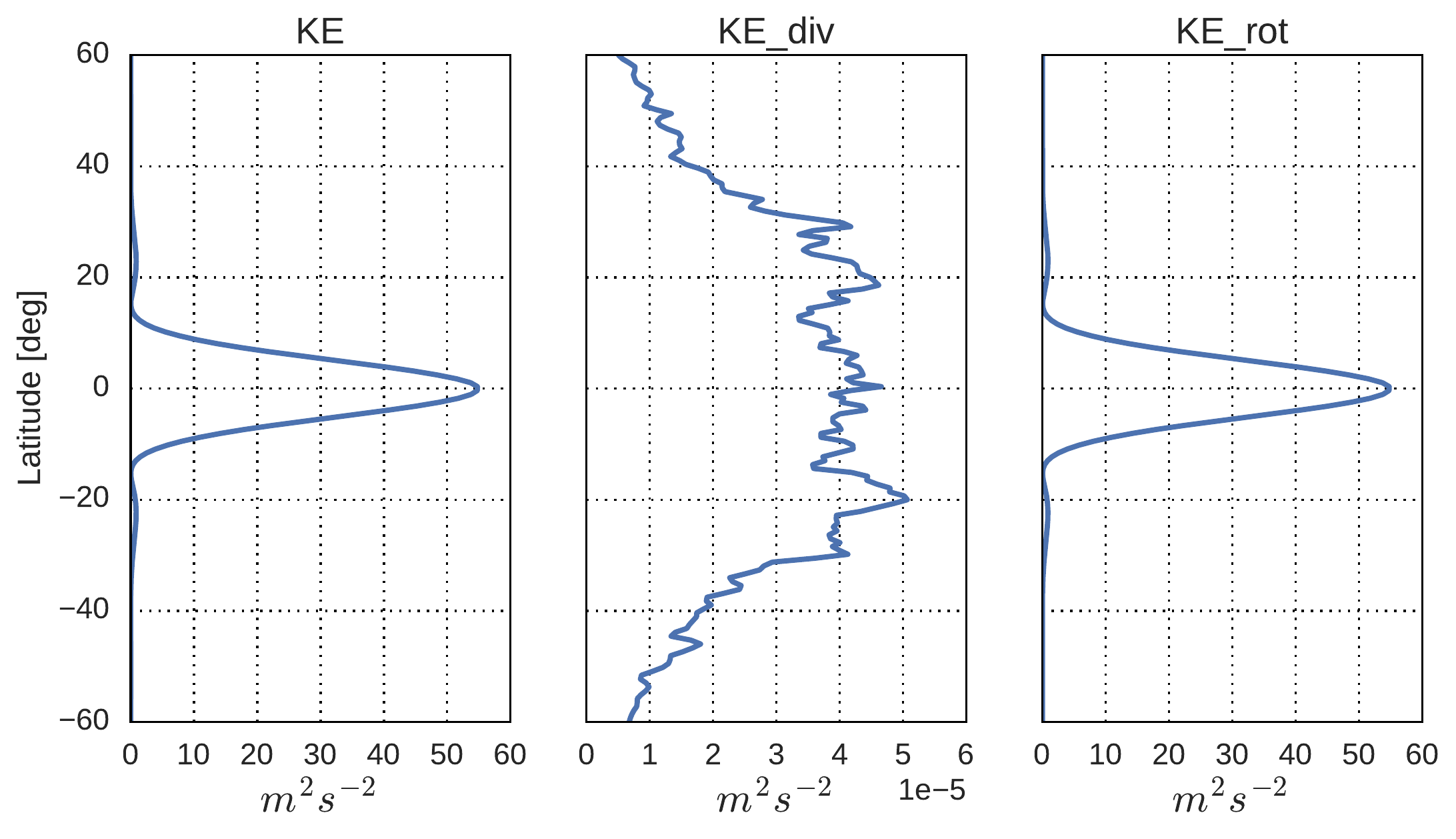}
\caption{\label{fig_ke_turb} Zonal mean kinetic energy and its divergent and rotational components for small-scale random forcing (averaged over days 500-2000).}
\end{figure}

Note that here too, $\overline{F_u}=0$, indeed, we impose this at every time step in the numerical simulation. A snapshot of the statistically steady state 
obtained 
with this forcing protocol is shown in Figure \ref{fig_height_turb}. 
In contrast, to the stationary state with large-scale forcing, we clearly have a 
solution with a much richer spatial structure. 
In fact, Figure \ref{fig_spectra}, which shows the scaling of kinetic energy versus wavenumber, 
suggests a distribution of energy 
across all scales. Given that the forcing is restricted to $k=100$, this brings forth the role of interscale energy transfer by the nonlinear terms in the 
SWE. Interestingly, the kinetic energy scales with a near power-law and an exponent of -5/3. 
Once again, the rotational component of the 
flow is dominant as is seen from Figure \ref{fig_ke_turb}. Indeed, this dominance of the 
rotational component is consistent with an inverse energy transfer regime for the rotating SWE \citep{farge,yuan}. 
It is worth noting that along with this scaling, it is still possible to discern systematic highs and lows in Figure \ref{fig_height_turb}.

\begin{figure*}[]
\centering
\includegraphics[width=15cm]{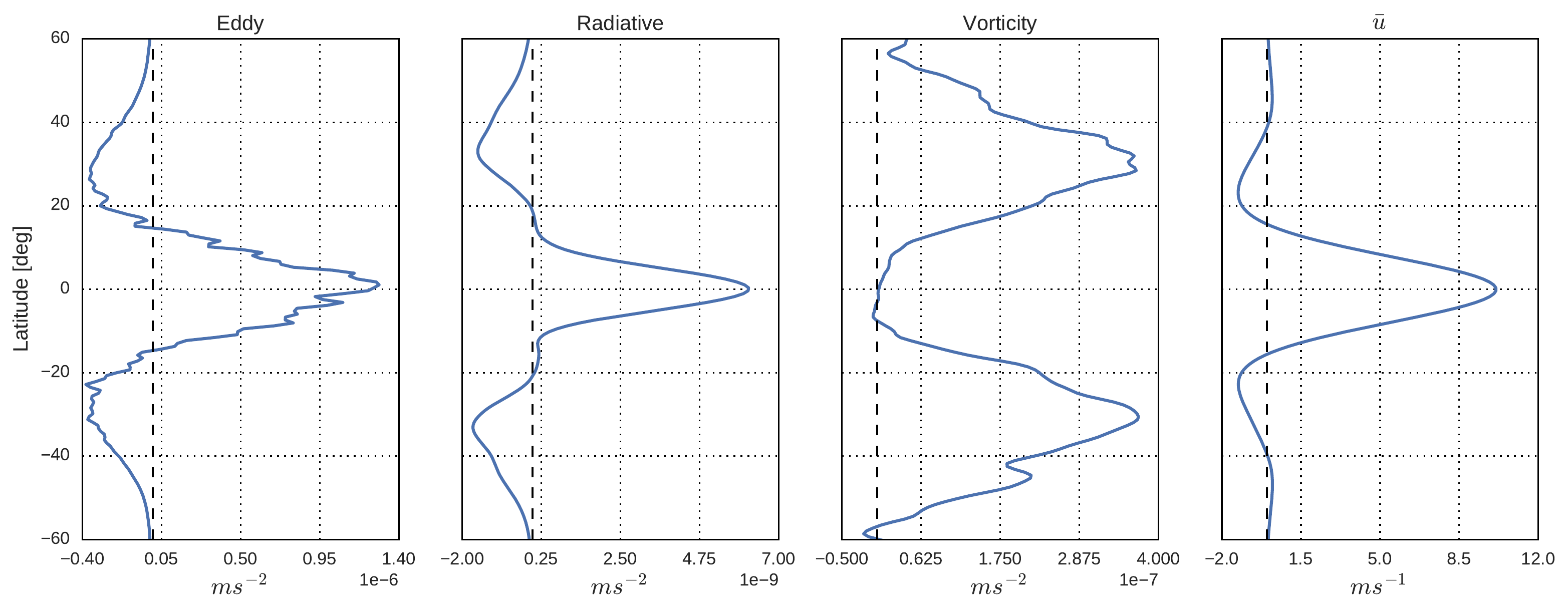}
\caption{\label{fig_flux_turb} Eddy, radiative and vorticity fluxes along with $\overline{u}$ for small scale random forcing (averaged over days 500-2000). Black dashed line denotes the zero flux.}
\end{figure*}

The momentum fluxes and the zonal mean zonal flow corresponding to this small-scale random forcing run are shown in Figure \ref{fig_flux_turb} (the components that are not shown are more than an order of magnitude smaller than the eddy and vorticity fluxes). We immediately see that the system settles down into a state of persistent
SR. Note that, while the radiative flux also appears
to provide an eastward acceleration, it is almost two orders of magnitude smaller than the eddy fluxes. Thus, once again, the eddy flux is the primary driver
of equatorial SR. Finally, returning to (\ref{a2}), when averaged over a long period, we again find a fairly close balance between $\overline{u}/\tau_m$ and 
$\overline{v' \omega'}$ at the equator. 

\section{Summary and Discussion}

In this work we have examined SR states in the spherical SWE. At the outset, working with a slightly stricter notion of SR (i.e., a positive
maximum of $\overline{u}$ at the equator), and closely following \cite{shp1}, it is seen that such a state can be supported either by direct forcing
(via $\overline{F_u}$) or by an eddy flux (via the correlation $\overline{v'\omega'}$). Indeed, \cite{shp1} have themselves provided an example of robust SR in the SWE
by means of direct forcing of the zonal momentum equations\footnote{As it happens, if one uses a traditional notion of SR, i.e., westerlies at the equator,
then a single, positive, off-equatorial vortex produces the desired result.}. Here, we try to construct simple examples wherein SR emerges with the 
constraint that $\overline{F_u}=0$ at the equator, in other words, the required eastward acceleration is provided by the eddy fluxes.

In the first example, we focused on large-scale vorticity forcing. Specifically, two equal and oppositely signed vortices were
placed on the equator. With continual forcing, the stationary solution had a strong 
subtropical signature consisting of anticyclonic and cyclonic pairs that formed a quadrupole. 
Further, it in the presence of both radiative damping and drag, eddy fluxes dominated the equatorial momentum budget and 
led to SR. The sensitivity of the solution to the radiative damping and
momentum drag timescales was also examined. It was seen that the tilt of the eddies (in a NW-SE direction) was largest for low damping and drag, 
and the solution was 
progressively more disconnected from the tropics with a decreasing strength of the damping and drag terms. We also examined the stationary 
solution to individual vortices, here the system does not show SR (at least not in our slightly stricter sense, though the winds at the equator are
westerly in nature), rather the zonal mean zonal flow is dominated by oppositely signed jets located off the equator. Comparing the fluxes
with these individual vortices to the solution with twin forcing, a fair degree of linearity was noted in the response for these
forcing protocols\footnote{This linearity can be exploited to ``guess" a forcing that would be needed for a particular $\overline{u}$. For example,
a mass sink along with the twin vortices at the equator leads to a double-horned profile much like that on (equatorial) Jupiter.}. 

After studying the stationary solution, we proceeded to examine the initial value problem with the twin, oppositely signed vortices. Here too, a 
subtropical quadrupole structure was established quickly, and persisted throughout the simulation when the drag and damping were 
switched off. Moreover, the solution was characterized by SR, again driven by the eddy fluxes. Interestingly, the solution obtained showed
a spontaneous westward propagation and circumnavigated the globe. Further, on examining the height fields through time,
the quadrupole continued to be well defined. This indicates that the solution to the initial value problem retained its coherence over 
a long period of time. Notably the initial value problem exhibits SR even in the absence of both drag and
damping, i.e., with no large-scale dissipation.

It is worth emphasizing that one of the main differences between our work and that of \cite{si} and \cite{wd} --- who built on the 
observation of \cite{andrew} 
that different dissipation mechanisms 
can affect the eddy fluxes in very different ways --- is that their results show SR only with radiative damping. In fact, in their 
simulations, 
SR disappears when momentum drag is switched on. In our case, the stationary solution exhibits SR in the presence of both 
radiative damping and momentum drag. Further, the initial value problem shows SR even in the 
absence of any large-scale damping mechanism. 

Finally, we looked at the effects of small-scale random forcing. At every step in the numerical scheme it was ensured that $\overline{F_u}=0$. 
Here too, SR set in and was supported by the eddy momentum fluxes. Also, in contrast to global random small-scale forcing
(where SR is established with only radiative damping),
the SR state emerged in the presence of both
momentum drag and radiative damping. Quite naturally,
the solution with random forcing was turbulent in nature, specifically, the power spectrum of the statistically steady state showed energy at all scales 
and followed a power-law with an approximate -5/3 exponent. Given that the forcing was restricted to very small scales, this immediately indicates the 
nonlinear interscale transfer of energy during the formation of a statistically steady state. Thus, in contrast to the linearity seen in the 
stationary solutions to large-scale forcing, here the SR state emerges 
from dynamics that are quite nonlinear in nature. But, in both examples, the rotational eddy flux, i.e., the correlation $\overline{v' \omega'}$, 
provides
the acceleration required to sustain a positive maximum of the zonal mean zonal flow at the equator.

{\it Acknowledgments: The authors would like to thank Prof. Dunkerton  for his valuable comments and suggestions which greatly helped in improving the quality of this paper.
Further the authors would like to thank the Divecha Centre for Climate Change at the
Indian Institute of Science for financial support.
JS would like to acknowledge the support from  IITM/MAS/DSG/0001 under Monsoon Mission funded by the Ministry of Earth Sciences, India.
JMM is supported by the Swedish E-science
Research Centre.}

\end{document}